\documentclass[showpacs,superscriptaddress,twocolumn,a4paper]{revtex4}

\usepackage{bm}
\usepackage{graphicx}
\usepackage{amsmath,amssymb}
\usepackage{hyperref}
\usepackage[latin1]{inputenc}

\begin{document}

\newcommand{\sign}{\operatorname{sign}}
\newcommand{\Ci}{\operatorname{Ci}}
\newcommand{\Si}{\operatorname{Si}}
\newcommand{\tr}{\operatorname{tr}}

\newcommand{\beq}{\begin{equation}}
\newcommand{\eeq}{\end{equation}}
\newcommand{\beqn}{\begin{eqnarray}}
\newcommand{\eeqn}{\end{eqnarray}}

\newcommand{\slp}{\raise.15ex\hbox{$/$}\kern-.57em\hbox{$ \partial $}}
\newcommand{\lnA}{\raise.15ex\hbox{$/$}\kern-.57em\hbox{$A$}}
\newcommand{\unmedio}{{\scriptstyle\frac{1}{2}}}
\newcommand{\uncuarto}{{\scriptstyle\frac{1}{4}}}

\newcommand{\trial}{_{\text{trial}}}
\newcommand{\true}{_{\text{true}}}
\newcommand{\const}{\text{const}}

\newcommand{\intp}{\int\frac{d^2p}{(2\pi)^2}\,}
\newcommand{\intx}{\int_C d^2x\,}
\newcommand{\inty}{\int_C d^2y\,}
\newcommand{\intxy}{\int_C d^2x\,d^2y\,}

\newcommand{\bP}{\bar{\Psi}}
\newcommand{\bc}{\bar{\chi}}
\newcommand{\hs}{\hspace*{0.6cm}}

\newcommand{\bra}{\left\langle}
\newcommand{\ket}{\right\rangle}
\newcommand{\bracket}{\left\langle\,\right\rangle}

\newcommand{\D}

\newcommand{\N}{\mbox{$\mathcal{N}$}}
\newcommand{\Lag}{\mbox{$\mathcal{L}$}}
\newcommand{\V}{\mbox{$\mathcal{V}$}}
\newcommand{\Z}{\mbox{$\mathcal{Z}$}}
\newcommand{\A}{\mbox{$\mathcal{A}$}}
\newcommand{\B}{\mbox{$\mathcal{B}$}}
\newcommand{\C}{\mbox{$\mathcal{C}$}}
\newcommand{\E}{\mbox{$\mathcal{E}$}}


\title{Pumping current of a Luttinger liquid with finite length}
\author{Sebasti\'{a}n Franchino Vi\~{n}as}
\affiliation{Departamento de F\'{\i}sica, Facultad de Ciencias
Exactas, Universidad Nacional de La Plata and IFLP-CONICET, CC 67,
 1900 La Plata, Argentina.}
\affiliation{Dipartimento di Fisica, Universit\`a di Roma "La
  Sapienza" and INFN, Sezione di Roma1, Piazzale Aldo Moro 2,
  I-00185 Roma, Italy}
\author{Pablo Pisani}
\affiliation{Departamento de F\'{\i}sica, Facultad de Ciencias
Exactas, Universidad Nacional de La Plata and IFLP-CONICET, CC 67,
 1900 La Plata, Argentina.}
\author{Mariano Salvay}
\affiliation{Departamento de F\'{\i}sica, Facultad de Ciencias
Exactas, Universidad Nacional de La Plata and IFLP-CONICET, CC 67,
 1900 La Plata, Argentina.}

\begin{abstract}We study transport properties  in a Tomonaga-Luttinger
liquid in the presence of two time-dependent point like weak
impurities, taking into account finite-length effects. By employing  analytical methods and performing a perturbation theory, we compute the backscattering pumping current ($I_{bs}$) in different regimes which can be established in relation to the oscillatory frequency of the impurities and to the frequency related to the length and the renormalized velocity (by the  electron-electron interactions) of the charge density modes.  We
investigate the role played by the spatial position of the impurity potentials.  We
also show how
the previous infinite length results for $I_{bs}$
are modified by the finite size of the system.
\end{abstract}\pacs{71.10.Pm,  73.63.Nm, 05.30.Fk, 72.10.Bg, 72.10.Fk} \maketitle

\section{Introduction}

In recent years there has been an intense focus on the analysis of quantum transport and non-equilibrium situations in the context of electrons in low dimensionality, such as quantum wires and carbon nanotubes \cite{Chamon0,Giamarchi, Ventra, Nazarov}.  In particular, the problem of electronic transport through a time-dependent perturbation has been studied in relation to the X-ray excitation \cite{Gogolin1} and the possibility of charge and spin exchange on conductors and semiconductors \cite{Chamon2}.  The investigation of the role of dynamic sources in highly correlated electron systems in 1D reveals an interesting equivalence with quantum evaporation of helium superfluids experiments \cite{Nature}.  Possible experimental realizations are a pump laser applied on a carbon nanotube producing a periodic deformation in the network structure that can be understood as an effective time-dependent impurity \cite{Science, Science2} or a Hall bar with a constriction \cite{Miliken}.  A detailed knowledge of quantum wires behavior in the presence of time-dependent perturbations will facilitate the development of devices based on quantum computation, single electron transport and quantum interferometers \cite{Fujisawa, Torres}.

In the theoretical  study of dynamic impurities in Luttinger liquids, an observable of special interest is the dc component of the backscattered current $I_{bs}$, which measures the rate of change of the total
number of right (or left) movers in the system due to the backscattering
impurities \cite{Chamon 1}.  For a point like time-dependent oscillatory impurity,  $I_{bs}$ has the same sign as the background current (proportional to the external voltage)  for strong repulsive interaction and as a consequence the conductance of a one-channel quantum wire grows \cite{Feldman, Schmeltzer}.  When a local barrier is switched on at finite time, the backscattered current decays with time in a way that crucially depends on electron-electron interactions \cite{yo 1}.  Another interesting problem is the behavior of the current when  switching processes in the interaction between the wire and the contacts are taken into account \cite{Perfetto}.

The presence of several oscillatory impurities produces another interesting effect in low dimensional systems: a pumping current, i.e. the persistence of $I_{bs}$  even in the absence of external voltage. These systems can be interpreted as
rectifiers, since they are characterized by the induction of a directed current with pure ac driving \cite{pump 0}.  In recent years there have been experimental observations of the quantum pumping effect, including the periodic deformation of the walls of quantum dots \cite{exp 2, exp 3, exp 4, exp 5} and the induction of currents by applying surface acoustic waves in carbon nanotubes \cite{exp 6}.  The pumping current was studied in one-dimensional systems of  non-interacting electrons with different geometries like wires or rings \cite{pump 2, pump 3, pump 5, pump 6, pump 7}, quantum dots \cite{pump 9,pump 10, pump 11} and in a  quasi one-dimensional graphene ribbon \cite{pump 4,pump 8}.
In the context of Luttinger liquids, the pumping current was computed for infinite length  at zero and finite temperature \cite{theories, yo 2}, where a power
(exponential)-law dependence with the frequency, the spatial separation between the impurities and the temperature was found in different energy regimes with exponents which are functions of the electron-electron interactions.  We emphasize that all these results were obtained taking as irrelevant the energy scale associated with the size of the wire.  A computation of the pumping current in Luttinger liquids  taking into account finite length does not seem to have been made before.  It is then very important, in order to make more-realistic predictions, to understand the role that
length plays in the pumping current.  This is the main purpose of this article.  It is worth mentioning that the effects of finite length
on a quantum wire transport properties were previously investigated for the
cases of a static impurity  \cite{Dolcini, Dolcini2} and a single dynamic impurity \cite{yo 3, Cheng}.

In this work, we study the pumping current in a Tomonoga-Luttinger liquid with finite length.  By performing a perturbative expansion in the backscattering amplitude and using a real time formalism \cite{Keldysh}, we obtain an analytical expression for $I_{bs}$.  We will restrict our analysis to the zero temperature limit.   The paper is organized as follows. In Section II we present the
model and  recall the results obtained for an infinite wire. Section III contains the original contributions of this paper. We present an analytical computation of the backscattered pumping current taking into account  the effect of
the finite length $L$.  Then, we examine the results in different regimes which can be established as function of the quotient of the frequencies associated to the oscillatory impurities and the length.  We also study the role  of the position of the barriers.  Finally, in Section IV, we summarize our results and conclusions.

\begin{figure}
\includegraphics{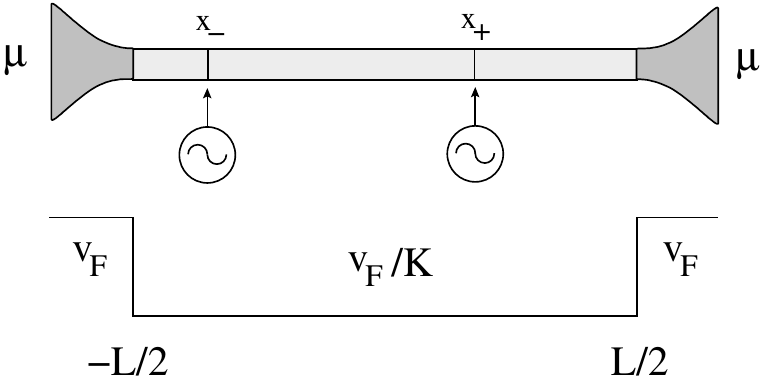}
\caption{\label{fig0ne}: The figure shows a quantum wire of length $L$ and Luttinger parameter  $K$  coupled adiabatically to two electrodes (Fermi systems with velocity $v_{F}$) with the same chemical potentials.  In the wire there are two backscattering oscillating impurities whit the same frequency.}
\end{figure}

\section{The model and review  of results at infinite length}
We consider a system of interacting electrons in a one dimensional space, composed by a clean quantum wire of length $L$ adiabatically coupled to two noninteracting electrodes at its end points $x = \pm L/2$.    We restrict our study to the case of spinless electrons and zero temperature. We will describe this system using the Tomonaga-Luttinger model, which represents fermions with a linearized dispersion relation and with a local forward-scattering interaction between them.  The Hamiltonian then reads
\begin{multline} H = \int^{\infty}_{- \infty} d x \Bigg\lbrace - i \hbar v_{F} \left[ \psi^{\dag}_{R}\partial_{x} \psi_{R} - \psi^{\dag}_{L}\partial_{x} \psi_{L} \right] \\ + g(x)\left[\psi^{\dag}_{R} \psi_{R} + \psi^{\dag}_{L} \psi_{L}\right]^{2} + \mu(x)\left[\psi^{\dag}_{R} \psi_{R} + \psi^{\dag}_{L} \psi_{L}\right]\Bigg\rbrace + H_{imp} ,\end{multline}where $\psi_{R}$ and $\psi_{L}$ are the fermionic field operators of the right- and left-moving electrons (we have omitted for the sake of simplicity their spatial-temporal dependence) and $v_{F}$ is the Fermi velocity. The function $g(x)$ describes the electron-electron interaction and its value is a constant $g$ in the bulk of the wire and zero in the bulk of the reservoirs; it is considered to change smoothly from $0$ to $g$ at the contacts within  a length of scale $d  \gg \Lambda$, where $\Lambda$ is a short-distance cutoff associated to the Fermi wavelength ($\Lambda \sim \hbar/k_{F}$).  We also assume that $L \gg  d$, i.e. the wire has a well defined length $L$.

The additional term \begin{multline}\label{himp}  H_{imp} = g_{B}\sum_{\pm} \int^{\infty}_{- \infty} d x\,   \delta(x - x_{\pm}) \cos[\Omega
t + \delta_{\pm}]\\ \times\left\{\psi^{\dag}_{R} \psi_{L}\exp[- \frac{2 i k_{F}x}{\hbar}] + \psi^{\dag}_{L} \psi_{R}\exp[\frac{2 i k_{F}x}{\hbar}]\right\},\end{multline}represents the interaction
of spinless electrons with two dynamical impurities located at points $x_{+}$ and $x_{-}$ in the wire (i.e $|x_{\pm}| < L/2$), with initial phases $\delta_{+}$ and
$\delta_{-}$ and oscillating both with frequency $\Omega$ and coupling
amplitude $g_{B}$.  In expression (\ref{himp}) we only take into account  backscattering between electrons and impurities because forward scattering does not change the transport properties here studied, at least not the lowest-order term of a perturbative expansion in the couplings.

The left and right electrodes are considered noninteracting electron reservoirs of semi-infinite length.  The function $\mu(x)$, which describes the chemical potential in the systems, is taken to be $\mu(|x| > L/2) = \mu$ and $\mu(|x| < L/2) = 0$, i.e. both electrodes have the same chemical potential so that the external voltage applied to the quantum wire (defined  as the difference between the chemical potentials in the leads) is zero.

The assumption that the variation of $g(x)$ is smooth in the contacts \cite{Dolcini, Dolcini2} and that the electrodes are held at the same temperature and their fermionic distributions are in equilibrium \cite{GGM} allows us to apply standard bosonization techniques
\cite{standardbos,makogon}: the fermionic operators are expressed in terms of the bosonic field $\Phi$ via the equation  $\psi^{\dag}_{R (L)} \psi_{L (R)}\simeq (2 \pi \Lambda)^{-1}\exp[+(-)2 i (\frac{ k_{F}x}{\hbar} + \sqrt{\pi
v_{F}}\Phi]$, while $\Phi$ is related to the charge density by the expression $\psi^{\dag}_{R}\psi_{R} + \psi^{\dag}_{L}\psi_{L} = (2 \pi v_{F} )^{1/2}\partial_x\Phi$.  Under the conditions described above, the Hamiltonian $H - H_{imp}$ in the bosonized language is quadratic
in the field $\Phi$.  After a shift to reabsorb the linear term in $\Phi$ associated to the chemical potential, we obtain the following effective
Lagrangian density:\begin{equation}\label{L0}\mathcal{L} = \frac{1}{2}
v(x)^{2}\left(\partial_{x}\Phi(x,t)\right)^2 -\frac{1}{2}
\left(\partial_{t}\Phi(x,t)\right)^2 \, + L_{imp} \,  , \end{equation}that describes
a spinless Tomonaga-Luttinger liquid with renormalized velocity  $v(x) = v = v_{F}/K$ when $|x| < L/2$ and $v(x) = v_{F}$ when  $|x| > L/2$. Here $K = 1/\sqrt{1 + 2 g /\pi \hbar v_{F}} $ measures the strength of the electron-electron interactions;  for
repulsive interactions $K < 1$, and for noninteracting electrons $K
= 1$.  The way $v(x)$ varies near the contacts from $v_{F}/K$ to $v_{F}$ is not physically relevant to our purposes because we take $d \ll L$. Therefore, we adopt for simplicity  the steplike function plotted in Figure \ref{fig0ne}.  $L_{imp}$ contains the impurities backscattering contribution in the wire:
\begin{multline} L_{imp}= - \frac{
g_{B}}{\pi \hbar \Lambda} \sum_{\pm} \delta(x - x_{\pm}) \cos[\Omega
t + \delta_{\pm}]\\ \times \cos[2 k_{F}x/\hbar + 2\sqrt{\pi
v_{F}}\Phi(x, t)] \, .\end{multline}

Since there is no external voltage, in the absence of impurities the background current is zero. On the other hand, in the presence of  impurities a backscattered  pumping current $I_{bs}$ appears and the total current (in the direction from the left electrode to the right electrode) is $I = - I_{bs}$.  The operator associated with the backscattered current, which measures the rate of change of the total
number of left or right movers in the wire due to the backscattering
impurities is defined as \cite{Feldman, theories, makogon}\begin{equation}\widehat{I}_{bs}(t) = e \frac{d N_{L}}{d t} = i e [H_{imp}, N_{L}]/\hbar = - e \frac{d N_{R}}{d t}\, , \end{equation}where \begin{equation}N_{L(R)} = \int^{L/2}_{-L/2} d x \, \psi^{\dag}_{L(R)}\psi_{L(R)}\end{equation}represents the total number of left- (right-) moving electrons in the wire.  In terms of the bosonic field $\Phi$, the backscattered operator is given by the equation \begin{multline}\widehat{I}_{bs}(t) = \frac{g_{B} e}{\pi \hbar \Lambda} \sum_{\pm}  \cos[\Omega t
+ \delta_{\pm}]  \\ \times \sin[2 k_{F}x_{\pm}/\hbar + 2\sqrt{\pi
v_{F}}\widehat{\Phi}(x_{\pm}, t)] \, .\end{multline}

Notice that $\widehat{I}_{bs}(t)$ is, by
definition, independent of the position on the wire $x$. Indeed,
since it is connected with the time evolution of the total number of left or
right moving particles, it involves an integral of the corresponding
density over the $x$-variable. Of course, it does depend on the
positions $x_{\pm}$ of the impurities in the wire.  The backscattered current at any time t is given by\beq  I_{bs}(t)  = \langle 0|S(- \infty ;
t)\widehat{I}_{bs}(t)S(t ; - \infty) | 0 \rangle \, ,  \label{uno} \eeq where $ \langle 0|$
denote the initial state and $S$ is the scattering matrix, which to the lowest order in the coupling $g_{B}$ is given by\beq S(t ; - \infty) = 1 - i \int^{\infty}_{- \infty} d x \int^{t}_{-
\infty} L_{imp} (t') d t' \, . \label{dosa}\eeq

When one inserts (\ref{dosa}) into (\ref{uno}) one finds several terms
of the form\beq A_{\alpha,\beta}=\langle 0|\exp[ 2 i \alpha \sqrt{\pi v_{F}}
\widehat{\Phi}(x', t')]\exp[- 2 i \beta \sqrt{\pi v_{F}}
\widehat{\Phi}(x, t)] |0 \rangle, \eeq with $\alpha,\beta=\pm 1$.
This kind of vacuum expectation values (v.e.v.) of vertex operators
has been computed many times in the literature. It is well-known
that $A_{\alpha,-\alpha}=0$ and thus the building block of our
computation is $A_{1,1} = A_{-1,-1}$.   Using Baker-Campbell-Hausdorff formula and the Debye-Waller general relation\cite{Giamarchi}, $A_{1,1}$  can be written as the exponential of a v.e.v.:\begin{equation}A_{1,1} = \exp\big[4 \pi  v_{F} \,
\langle 0|\widehat{\Phi}(x', t')\widehat{\Phi}(x, t) - \frac{\widehat{\Phi}^{2}(x', t') + \widehat{\Phi}^{2}(x, t)}{2} | 0 \rangle
\big].\label{factorA}\end{equation}

Now, we review the result for the pumping current in the case of $L = \infty$ \cite{theories}.  In order to explicitly evaluate the previous expressions we
need Keldysh \cite{Keldysh} lesser function $G^{<}$ at infinite length and zero temperature given by\begin{multline} \langle 0|\widehat{\Phi}(x', t')\,\widehat{\Phi}(x, t)| 0
\rangle = i
G^{<}(x,t;x',t') \\ = \frac{1}{2\pi}\,\int\,dp\,d\omega\,e^{i p (x - x') - i \omega(t-t')}\,\theta(-\omega)\,\delta(\omega^2-v^2p^2).\end{multline}

Joining these last two equations we obtain\begin{equation} A_{1,1} =  \frac{ \Lambda ^{2 K}}{\big((x - x')^{2} - (v(t - t') + i\Lambda)^{2}\big)^{K}} .\label{tres1}\end{equation}

Then, using (\ref{tres1}) in the computation of (\ref{uno}), we
find the following expression for $I_{bs}$:\begin{equation} I_{bs} = \frac{g_{B}^{2} e \Lambda ^{2 K - 2} \sin[\frac{2 k_{F}
a}{\hbar}] \sin[\phi]}{4 \pi^{2} \hbar^{2} }\int^{\infty}_{-\infty}\frac{\exp[i \Omega t']\, d t'}{\big( a^{2} - (\Lambda + i v t')^{2}\big)^{K}}. \label{pump}\end{equation}
In this equation $ a = x_{+} - x_{-} $ represents the spatial separation between the two impurities and
$\phi = \delta_{+} - \delta_{-}$ is their phase difference.  We observe that (\ref{pump})  is time-independent, this is due to the absence of external voltage.  In general, when an external voltage $V$ is applied  to the wire, the backscattered current at second order in $g_{B}$ has two parts: one
independent of time (which we can identify with the dc current), and  another varying harmonically
with time, with frequency $2 \Omega$ (which we can identify with the ac current that does not contribute to any charge transfer and whose average over the period $2 \pi/\Omega$ is zero) \cite{theories, yo 2}.  For the pure pumping case studied here ($V = 0$), the calculation of (\ref{uno}) gives automatically a vanishing ac current.

Now let us go back to equation (\ref{pump}). Performing the integral and defining the dimensionless frequency $\omega_{a} = \Omega |a|/ v$ associated with the separation between barriers,  $I_{bs}$ can be expressed as
\begin{equation} I_{bs} = \frac{e g_{B}^{2}\Lambda^{2 K - 2}\Omega^{2 K - 1} \sin[\frac{2 k_{F}
a}{\hbar}] \sin[\phi]}{2^{K + 1/2}  \sqrt{\pi}  \hbar^{2} v^{2 K} \Gamma[K]} \omega_{a} ^{1/2 - K}\, J_{K - 1/2}[\omega_{a}], \label{pu}\end{equation} where $\Gamma$ is the Gamma function and $J$ is the Bessel function of the first kind.  The factors $\sin[\frac{2 k_{F}
a}{\hbar}]$ and $\sin[\phi]$ are characteristic of a pumping current in one-dimensional systems, and show that the direction of $I_{bs}$ at zero voltage is determined by the spatial separation and phase difference between impurities.  The behavior  $I_{bs} \propto g_{B}^{2} \sin[\phi]$ was predicted for a two-barrier quantum pump
in a linear setup \cite{pump 9}, in an annular setup \cite{pump 2, pump 3} and  also obtained within Floquet scattering matrix
formalism \cite{pump 7}.  The proportionality $I_{bs} \propto \sin[\frac{2 k_{F}
a}{\hbar}]$ exclude the possibility of pure pumping with a single impurity ($a = 0$) in a quantum wire. Moreover, since
$\hbar/k_{F} \sim \Lambda \approx 1{\text{\AA}}$, the factor  $\sin[\frac{2 k_{F}
a}{\hbar}]$  is rapidly oscillating as a function of $a$.

The dimensionless parameter $\omega_{a}$  measures the relation between the separation of the barriers and the dynamic  length scale $v/\Omega$ associated to the frequencies of the impurities.  Large (small) values of  $\omega_{a}$ correspond to a scale regime of long (short) separation of the barriers.    From (\ref{pu}), we obtain as a result that in the scale regime of large separation   ($\omega_{a} \gg 1$), the pumping current goes as $\omega_{a} ^{- K} \cos[\omega_{a} - K \pi/2]$. We thus have a damped oscillatory function of $\omega_{a}$ with period $2 \pi$ and there is a suppression of $I_{bs}$ when $\omega_{a} = \frac{\pi}{2}(2 n + 1 + K)$ with $n$ natural.
On the other hand, in the scale regime of short separation ($\omega_{a}  \ll 1$) the current is independent of $\omega_{a}$.

Note that for  $K < 1/2$, the pumping current becomes large when $\Omega$ decreases. Hence, the perturbative expansion in powers of $g_{B}$ breaks down when $\Omega \rightarrow 0$.  Using a scaling analysis we can estimate that this expansion is valid when $\frac{g_{B}}{\hbar v} (\frac{\Lambda \Omega}{v})^{K - 1} \ll 1$.  We remark that expression (\ref{pu}) does not include the case $\Omega = 0$, where the pumping current is also zero.  All these statements imply that the current must be a nonmonotonic function of $\Omega$.  In order to determine this function one has to go beyond the lowest-order perturbative results.

\begin{figure}
\includegraphics{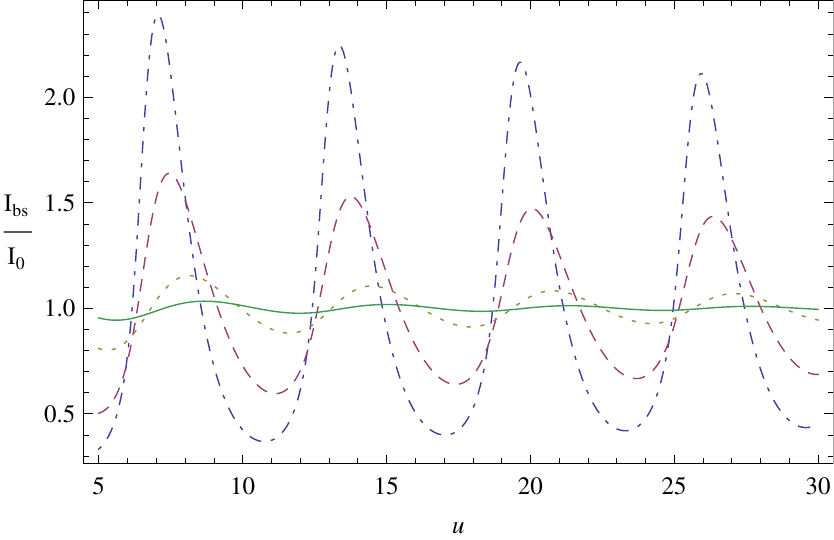}
\caption{\label{fig2ne1}: Pumping current at finite length ($I_{bs}$) divided by the pumping current at infinite length ($I_{0}$) as function of $u$ in the regimes of $\omega_{a}, \omega_{r} \ll 1$.  Dotdashed line corresponds to $K = 0.2$, dashed line to $K = 0.3$, dotted line to $K = 0.5$ and solid line to $K = 0.7$.}
\end{figure}

\section{Results at finite length}

In this section we present the main results of this work. We will
compute the backscattered pumping current at finite length.  The effect of the quantum interference originated by
reflections at the ends of the wire (which are now at finite
distances from the barriers) modifies drastically the expression  of the
backscattered current. The crucial point is that when the wire has
finite length $L$ the function (\ref{factorA}) becomes dependent
on $L$ and, since there is no translational invariance, not only on $x - x'$ but on $x$ and $x'$ separately . In this case, after computing (\ref{factorA}) (see Appendix A and \cite{Dolcini}), inserting the result in (\ref{uno}) and performing the rescaling $t \rightarrow v t/L$, we obtain for the
backscattered current $I_{bs}$, the expression
\begin{equation}I_{bs} = \frac{g_{B}^{2} e   \sin[\frac{2 k_{F}
a}{\hbar}] \sin[\phi]}{(2 \pi)^{2} \hbar^{2} \Lambda ^{2} (v/L)}\int^{\infty}_{-\infty}\exp\left[i \frac{\Omega L}{v} t\right]\, F(t)\, d t,\ \label{pump6}\end{equation}where:
\begin{multline}\label{F(t)}F(t) = \exp\left\lbrace - K \ln \left[ \frac{ \big((\Lambda/L) + i t\big)^{2} + (a/L)^{2}} {(\Lambda/L)^{2}}\right] \right.\\  - K\sum_{n \in Z even \neq 0} \gamma^{|n|}\ln \left[ \frac{ \big((\Lambda/L) + i t \big)^{2} + (n + a/L)^{2}} {n^{2}} \right] \\ \left.- K\sum_{n \in Z odd } \gamma^{|n|}\ln \left[ \frac{ \big((\Lambda/L) + i t \big)^{2} + (n + r/L)^{2}} {(n + a/L + r/L)(n - a/L + r/L) } \right] \right\rbrace. \end{multline}
In the last expression $\gamma = (1 - K)/(1 + K)$ is the Andreev-like reflection
parameter and  $r = x_{+} + x_{-}$ is the center of mass of the impurities. Notice from eq.\ (\ref{pump6}) that, as for the infinite wire, the current vanishes for the special case of a single impurity ($a=0$). This is also a consequence of the fact that, for a single impurity, the current depends on the position of the impurity only through the Keldysh lesser function $G^<$ (this is straightforward from eq. (\ref{uno})) which is invariant under spatial inversion. However, since the current changes sign under this transformation, i.e.\ under interchange of left and right, then it must be zero (this argument holds, of course, in the absence of an external voltage).

For $u = \frac{\Omega L}{v} \gg 1$,  i.e. when the frequency $\Omega$ of the barriers  is much bigger than the ballistic frequency $v/L$ related to the length of the quantum wire, we have the regime of long wires.  In realistic systems the renormalized velocity is in the range of $10^{6} \text{m}/\text{s}$, so that the wire length scale corresponds to a value of $L$ subject to the condition $L \gg  (10^{6} \text{m}/\text{s}) / \Omega$.   For example, for frequencies of order $10 ^{12} Hz$ and $10^{9} Hz$, the corresponding wire lengths  satisfy $L \gg  10^{-6} m$
and $L \gg  10^{-3} m$, respectively.  For carbon nanotube experiments a typical length is in the range that runs from 1 $\mu \text{m}$ to a few $\text{cm}$; hence,
values of $u >1$ should be observable in the high range of frequencies.  In this regime it is
possible to obtain an analytic expression for (\ref{pump6})
along the lines explained in Ref.
\cite{ponomarenko}.

In order to compute the integral in (\ref{pump6}), it is important to notice that the logarithms in (\ref{F(t)}) must be defined with their branch cuts in the negative real semi-axis. As a consequence, in the complex $t$-plane the function $F(t)$ has an infinite number of vertical branch cuts that go from $(n\pm x/L+i\Lambda/L)$ to $(n\pm x/L+i\infty)$ for each integer $n$ ($x$ refers to $a$ or $r$ depending on whether $n$ is even or odd, respectively). Therefore, the integral in (\ref{pump6}) amounts to infinitely many integrals in the complex $t$-plane along curves that enclose each of these vertical branch cuts. At this point, the rescaling $t\rightarrow t/u$ allows us to identify the leading contribution to $I_{bs}$ for large $u$. The integrals
around the cuts can be explicitly solved to leading order in $u$ and give the following result for $I_{bs}$:\begin{widetext}\begin{multline}I_{bs} = \frac{e g_{B}^{2}\Lambda^{2 K - 2} \Omega^{2 K - 1}\sin[\frac{2 k_{F}
a}{\hbar}] \sin[\phi]}{2^{K + 1/2}  \sqrt{\pi}  \hbar^{2} v^{2 K}} \Big\{D_{0}\,\omega_{a} ^{1/2 - K}\, J_{K - 1/2}[\omega_{a}] \\ - \sum_{n\in Z even\neq 0}D_{n}^{e}\,\omega_{a}^{1/2 - K\gamma^{|n|}}\, u^{2 K(\gamma^{|n|} - 1)} \left(\cos[|n| u + \pi \gamma^{|n|}
 ]\,J_{K\gamma^{|n|} - 1/2}[\omega_{a}]  - sgn[n a]\sin[|n| u + \pi \gamma^{|n|}]N_{K\gamma^{|n|} - 1/2}[\omega_{a}]\right) \\ - \sum_{n\in Z odd}D_{n}^{o}\,\omega_{r}^{1/2 - K\gamma^{|n|}}\, u^{2 K(\gamma^{|n|} - 1)}\left(\cos[|n| u + \pi \gamma^{|n|}]\,J_{K\gamma^{|n|} - 1/2}[\omega_{r}] -  sgn[n r]\sin[|n| u + \pi \gamma^{|n|}]N_{K\gamma^{|n|} - 1/2}[\omega_{r}]\right)\Big\}\label{pri}\end{multline}\end{widetext} where $\omega_{r} = \Omega |r|/ v$ , $sgn$ is the sign function, $N$ is the Bessel function of second kind and we have defined the following coefficients that depend on the location of the impurities and the electron-electron interaction:
\begin{equation} D_{0} = \prod_{m\in even \neq 0}\left|\frac{m+2 \frac{{a}}{L}}{m}\right|^{-K\,\gamma^{|m|}}\frac{1}{\Gamma[K]} ,\end{equation}
\begin{multline} D_{n}^{e} = \frac{|n (n+2 \frac{{a}}{L})|^{-K}}{2^{-K}\Gamma[K\gamma^{|n|}]}\left|\frac{n^3}{8(\frac{{a}}{L}+n)}\right|^{K\,\gamma^{|n|}}\\ \times\prod_{m\in Z even \neq \pm n,0}\left|\frac{m^2}{m^2-n^2+2\frac{{a}}{L}(m-n)}\right|^{K \gamma^{|m|}}
\\ \times \prod_{m\in Z odd }\left|\frac{(m+\frac{{r}}{L})^2-(\frac{{a}}{L})^2}{(m+\frac{{r}}{L})^2-(n+\frac{{a}}{L})^2}\right|^{K \gamma^{|m|}}\end{multline}
\begin{multline} D_{n}^{o} = \frac{|(\frac{{a}}{L})^2-(\frac{{r}}{L}+n)^2|^{-K}}{2^{-K}\Gamma[K\gamma^{|n|}]}\\ \times \left|\frac{(n^{2} - (\frac{{a}}{L} + \frac{{r}}{L})^{2})(n^{2} - (\frac{{a}}{L} - \frac{{r}}{L})^{2}) }{8(\frac{{r}}{L}+n)n}\right|^{K\,\gamma^{|n|}} \\ \times \prod_{m\in Z even }\left|\frac{m^2}{(m+\frac{{a}}{L})^2-(\frac{{r}}{L}+n)^2}\right|^{K\,\gamma^{|m|}}
\\ \times
\prod_{m\in Z odd \neq \pm n}\left|\frac{(m+\frac{{r}}{L})^2- (\frac{{a}}{L})^2}{m^2-n^2+2\frac{{r}}{L}(m-n)}\right|^{K\,\gamma^{|m|}}.\end{multline}
Result (\ref{pri}) is the  generalization to finite length (and considering $L \gg v/\Omega$) of the result shown in Section II for the pumping current.  Thus, we have obtained an analytical expression for $I_{bs}$ at the lowest-order in the impurity coupling $g_{B}$,  as a function of the length of the wire, the frequency and position of the impurities and the strength of the interaction between electrons. The pumping current at finite length  is a
superposition  of infinite damped oscillatory functions of $u$ with
period $2 \pi/|n|$, each of them decaying as $u^{2 K( \gamma^{|n|} -1)}$. The origin of the oscillation of
the pumping current in terms of the length is the interference effect of plasmon
modes which are reflected by both the impurities and the
wire-reservoir contacts.  In addition to  $\omega_{a}$ (defined in the previous section), the dimensionless parameter  $\omega_{r}$ characterizes the relation between the  medium position ($|r|$) of the impurities with the dynamic length scale $v/\Omega$, i.e. it is a measure of the symmetry of the impurities distribution in the wire.  Large (small) values of  $\omega_{r}$ correspond to a scale regime of low (high) symmetry  of the barriers position (with respect to the center of the wire).    Notice that in the case $L=\infty$, considered in the previous section, there was no dependence of $I_{bs}$ with the center of mass of the impurities. On the contrary, for finite $L$ there appears a dependence on $r$ due to the breaking of translational invariance.

\begin{figure}
\includegraphics{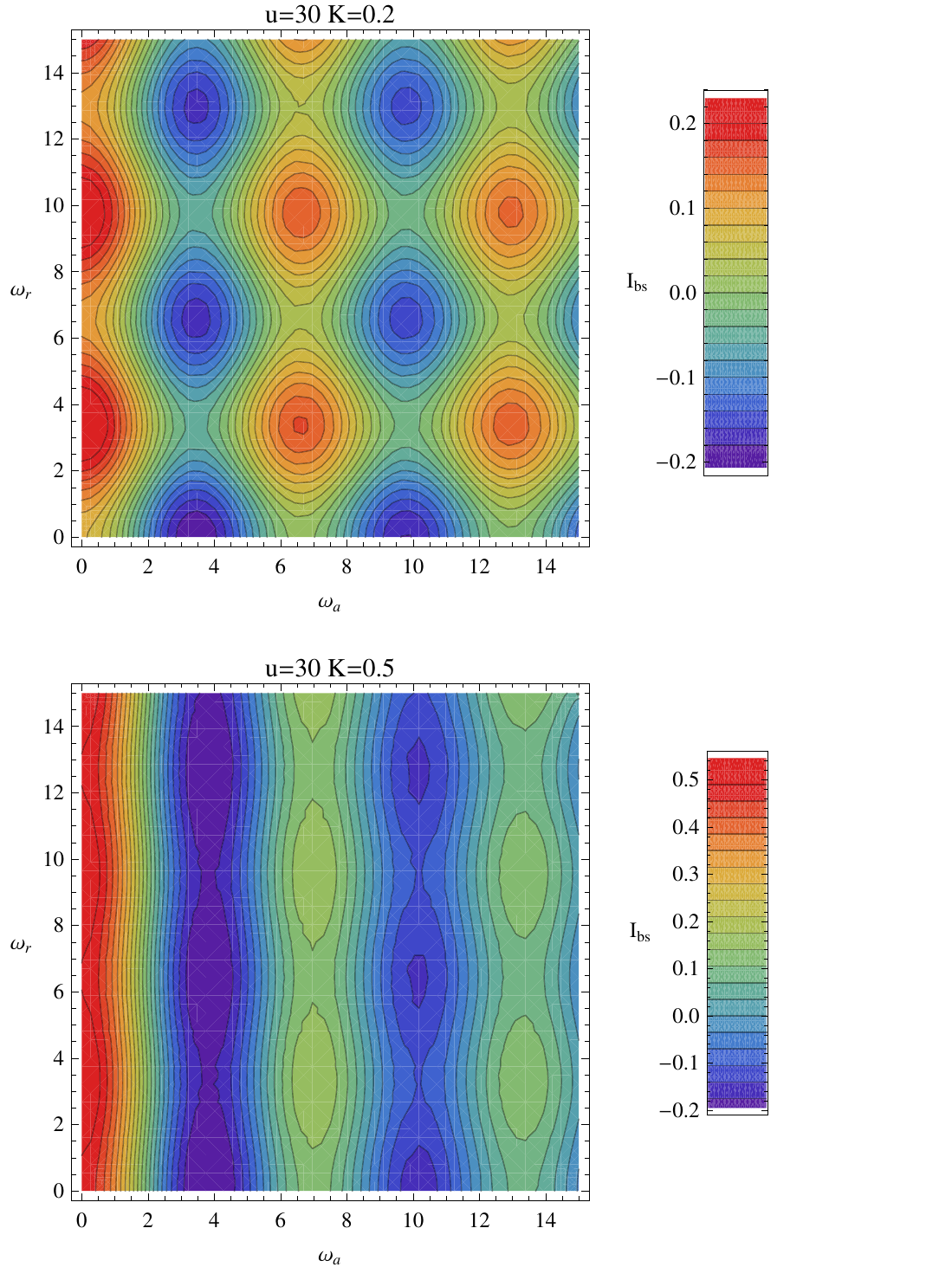}
\caption{\label{fig1ne1}(Color online) Pumping current as function of  $\omega_{a}$ and $\omega_{r}$ for two values of $K$. The units of $I_{bs}$ are set equal to $\frac{e g_{B}^{2}\Lambda^{
2 K - 2} \Omega^{2 K - 1}\sin[\frac{2 k_{F}
a}{\hbar}] \sin[\phi]}{2^{K + 1/2}  \sqrt{\pi}  \hbar^{2} v^{2K}}$ and we consider $a, r > 0$. }
\end{figure}
\begin{figure}
\includegraphics{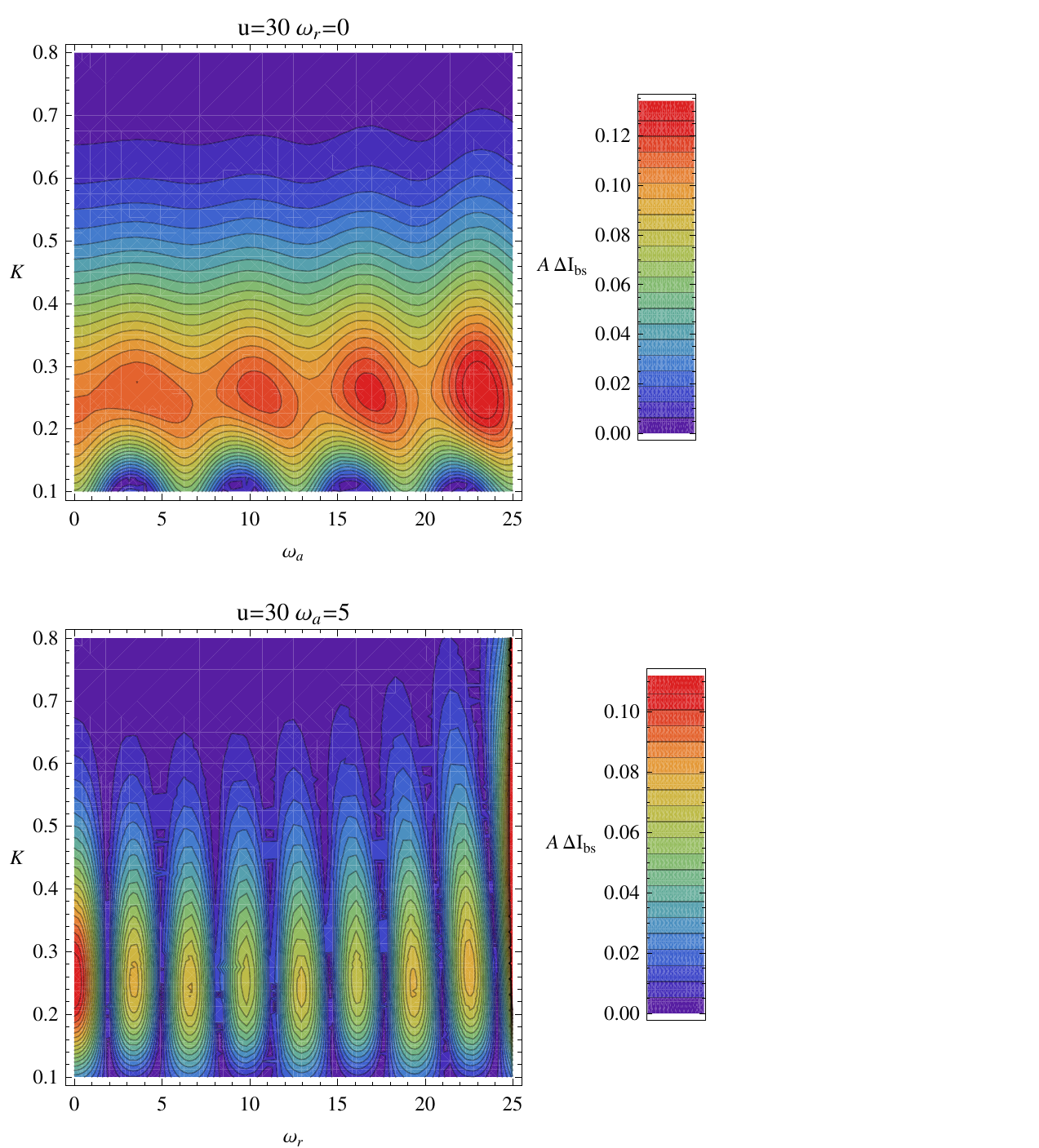}
\caption{\label{fig1ne2}(Color online) Absolute value of the difference between the pumping current at finite and infinite length  as function of $K$ and the geometry ($\omega_{a}$ or $\omega_{r}$).    The units of $I_{bs}$ is the same as in Fig. 3 and we consider $a, r > 0$.  }
\end{figure}

If we consider the limit $r\rightarrow 0$ (a symmetrical arrangement with respect to the center of the wire) and $a\rightarrow 0$ in equation (\ref{pri}) the pumping current acquires the compact form
\begin{multline}\label{eq:steps}
I_{bs} \approx I_{0}\Bigg\lbrace1-2\sum_{n=1}^{\infty}\prod^{\infty}_{m \neq n, m > 0} \left|\frac{m^{2}}{m^{2} - n^{2}}\right|^{2 K \gamma^{m}}\\ \times\frac{\Gamma(2K)2^{- 2 K \gamma^{n}}n^{2 K(\gamma^{n} - 1)}\cos[n u + \pi \gamma^{n}]}{\Gamma(2K\gamma^n)u^{2K(1-\gamma^{n})}}\Bigg\rbrace,
\end{multline}
where $I_{0}$ is the value for the current in the case $L = \infty$ computed in the previous section.  Expression  (\ref{eq:steps}) is a good approximation for the pumping current in the regimes $\omega_{r} \ll 1$ and  $\omega_{a} \ll 1$, that is when the center of mass and the spatial separation are much smaller than the length scale associated with the frequency  $v/\Omega$.  It is worth mentioning that in the case of noninteracting electrons ($K =1$ and $\gamma = 0$), denominators in (\ref{eq:steps}) go to infinity and the pumping current is equal to $I_{0}$, i.e., there is no finite length effect when $K = 1$.

In Figure \ref{fig2ne1} we have plotted (\ref{eq:steps}) as a function of $u$ and for different values of $K$.  The differences between $I_{0}$ and$I_{bs}$ are more pronounced for high electron-electron interactions ($K \rightarrow 0$).
Accordingly, in this case  the effect of the finite length remains important even for big values of $u$: the damping factor of each oscillatory function that represents a correction to $I_{0}$ has the form $C_{n} u^{2 K(\gamma^{n} - 1)}$, where $C_n$ is a coefficient determined by equation (\ref{eq:steps}).  Given that $C_{n + 1} <  C_{n}$ for any value of $K$, then the most important correction to $I_{bs}$ goes as $L^{2 K (\gamma - 1)}$.  For example, given $K = 0.25$, the dominant correction to $I_{0}$  for large $u$ is a term $\cos[u + 3 \pi/5]u^{-1/5}$ that shows the very slow convergence of the series.

We notice that this behavior is analog to the results previously obtained for the case of a static impurity in a finite quantum wire at non zero external voltage $V$ when the impurity is in the center of the wire \cite{Dolcini, Dolcini2} if we replace $\Omega$ by the Josephson frequency $e V/\hbar$ associated with the external voltage. That is to say, the distortion in the current with respect to the infinite length case corresponding to two oscillatory impurities with different positions (both close to the center of the wire) and at zero voltage is similar to the case of one static impurity at non zero voltage.

In the cases when $\omega_{r} \gg 1$ or  $\omega_{a} \gg 1$, we have a regime where the spatial distribution of the impurities with respect to the center of the wire and the spatial separation between them is greater than $v/\Omega$ (large tunneling).  Using the asymptotic expansion for the Bessel function for large values of their arguments, $N_{\alpha}(x) \sim \sin[x - \frac{\pi}{2}(\alpha + 1/2)]x^{-1/2}$ and $J_{\alpha}(x) \sim \cos[x - \frac{\pi}{2}(\alpha + 1/2)]x^{-1/2}$ for $x \gg 1$, we find that in addition to the dependence in $u$, expression (\ref{pri}) is a superposition of damped oscillatory functions in terms of  $\omega_{r}$ and $\omega_{a}$, each one with period $2 \pi$ and a decaying factor of the form $\omega_{a}^{- K \gamma^{2n}}$ and  $\omega_{r}^{- K \gamma^{2n + 1}}$ in the region given by $\omega_{a,r} \ll u$.  As an example, in Figure \ref{fig1ne1} we show the behavior of $I_{bs}$ as a function of the position of the barriers where this periodicity is manifest.
It is worth noting that the backscattering current for infinite $L$ is suppressed, as shown in the previous section, for particular values of $\omega_{a}$, namely $\omega_{a} \approx \frac{\pi}{2}(2 n + 1 + K)$. On the contrary, as can be seen from Figure \ref{fig1ne1}, for finite $L$ the regions on which $I_{bs}$ vanishes are determined by the values of $\omega_r$ as well.
Of course one can also notice that as the interactions decrease (larger $K$) the dependence on $\omega_r$ diminishes and the regions where $I_{bs}=0$ resemble
those of infinite $L$.

In Figure \ref{fig1ne2}  we show the  dependence of the absolute value of the  difference between the value of $I_{bs}$ at finite and infinite length with the Luttinger parameter $K$ and the relative and medium position of the impurities.
Although for weak interactions the effect of the length disappears, for high interactions it becomes relevant and the distortion with respect to the infinite length case is sensitive to the specific values of the length of the wire ($u$) and the position of the barriers.  In general, when $\omega_{a} + \omega_{r} \rightarrow u$ (that is, when one impurity is in any extreme of the wire) the difference with the infinite length case is more pronounced. One can see on the lower graph of Figure \ref{fig1ne2} that the same is true for a symmetric configuration ($\omega_{r} \rightarrow 0$).

We finally consider the regime of small length $u = \frac{\Omega L}{v} \ll 1$. The integral in expression (\ref{pump6}) is once more performed along the infinitely many branch cuts of $F(t)$ in the complex $t$-plane. The leading contributions to $I_{bs}$ for small $u$ are given by
\begin{multline}I_{bs} = \frac{e g_{B}^{2}\Lambda^{2 - 2 K} \Omega^{2 K - 1}\sin[\frac{2 k_{F}
a}{\hbar}] \sin[\phi]}{2 \pi \hbar^{2} v^{2 K}} \\ \times F_{0}\left(\frac{a}{L},\frac{r}{L}\right) u ^{2 - 2 K}\left[1 - u^{2}F_{1}\left(\frac{a}{L},\frac{r}{L}\right)\right], \label{chico}\end{multline}
where we have defined the factors
\begin{equation} F_{0} = \prod_{n\in Z even \neq 0} \left(n^{2}\right)^{K\,\gamma^{|n|}} \prod_{n\in Z odd} \left[\left(n + \frac{r}{L}\right)^{2} -
 \left(\frac{a}{L}\right)^{2}\right]^{K\,\gamma^{|n|}} \end{equation}and
\begin{multline} F_{1} = \frac{K}{3} \big[\sum_{n\in Z even }\gamma^{|n|}\left(1 + K \gamma^{|n|}\right)\left(2 n + \frac{a}{L}\right)^{2} \\ + \sum_{n\in Z odd }\gamma^{|n|}\left(1 + K \gamma^{|n|}\right)\left(2 n + \frac{r}{L}\right)^{2} \big].\end{multline}

\begin{figure}
\includegraphics{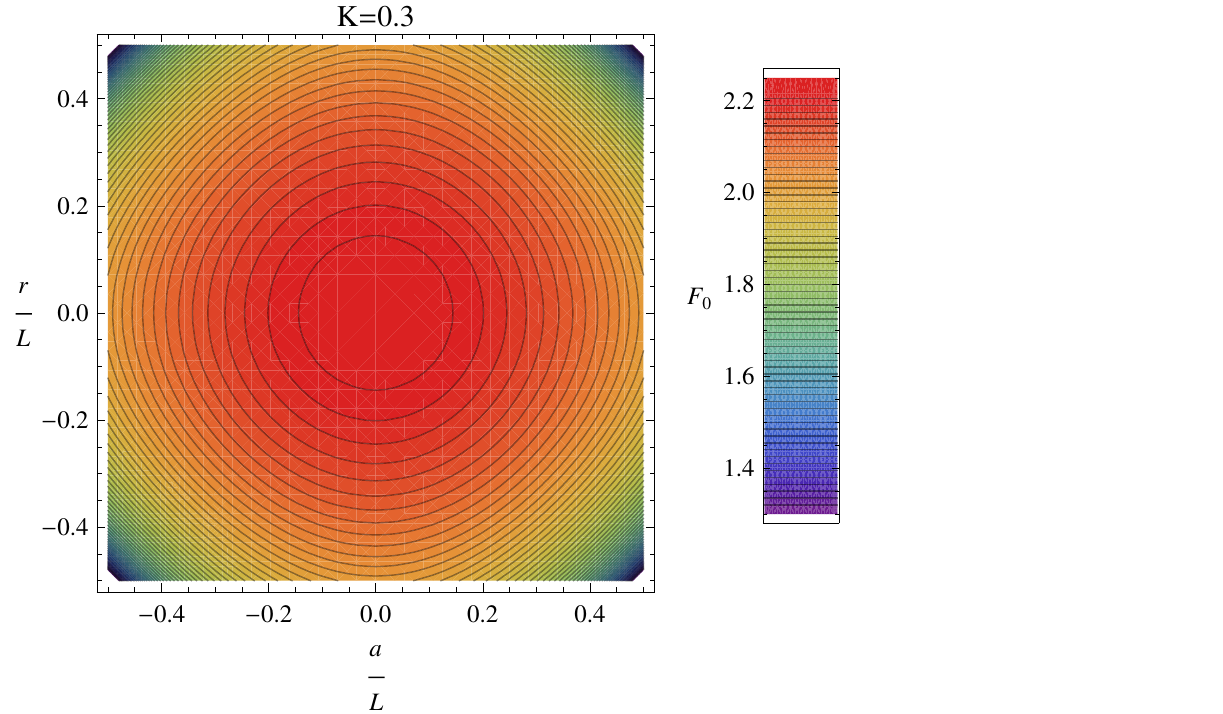}
\caption{\label{fig2ne2}(Color online) The factor $F_{0}$ describes the dependence with the geometry of $I_{bs}$ in the case of short wires.}
\end{figure}

Expression (\ref{chico}) shows that for short wires $I_{bs}$ goes as $u^{2 - 2K}$,
i.e. the current goes to zero when $L\rightarrow 0$; this suppression is more pronounced for stronger interactions between electrons.  When $K = 1$,
we have $F_{0} = 1$ and $F_{1} = 0$, i.e.\ the dependence with $L$ and the positions of the barriers is dropped and the current is the same to the case at infinite length with  noninteracting electrons.   In terms of the frequency and the length, the pumping current goes as $\Omega$ and $L^{2 - 2K}$ respectively.  We observe that, unlike the cases at infinite and large length, the collapse at $\Omega \approx 0$ disappears and then the pumping current goes to zero when $\Omega$ decreases.  In this case, the expansion in the coupling constant $g_{B}$ is valid when $\frac{g_{B}}{\hbar v} (\frac{\Lambda}{L})^{K - 1} \ll 1$.  Since the influence of the position of the barriers is dominated by the factor $F_{0}$,  in Figure \ref{fig2ne2} we plot  $F_{0}$ as a function of $r/L$ and $a/L$: it has a maximum for a symmetrical arrangement ($r =0$) and $a=0$ and decreases when $|a|$ or $|r|$ increase.

\section{Conclusions}
To summarize, we have analyzed the characteristics of time-dependent
transport in a Tomonaga-Luttinger liquid subject to a zero
bias voltage, when  two weak  barriers are oscillating in the wire.  We focused our attention on the
backscattered pumping current $I_{bs}$.

The novel
features of our investigation come from the consideration of
a wire of finite length $L$.  We analyzed the distortion of the pumping current with respect to the infinite length case.
In order to do so, we defined a dimensionless parameter $u = \Omega L/v$ and  presented an exact and analytical computation of $I_{bs}$ as a function of $u$ for long (short) wires, that is, when $u \gg (\ll) 1$.  For long wires the pumping current  is a
superposition  of infinite damped oscillatory functions of $u$ with
period $2 \pi/n$, with $n$ a positive integer, each of these oscillations has a decay
prefactor of the form $u^{2 K( \gamma^{n} -1)}$.  The origin of the oscillation of
the pumping current in terms of the length is the interference effect of plasmon
modes which are reflected by both the impurities and the
wire-reservoir contacts.
As expected, for strong electron-electron interactions ($K \rightarrow 0$) the distortion with respect to the infinite length case is more drastic and persists even for big values of $u = \Omega L/v$.  On the other hand, for weak interactions the effect of the length disappears and $I_{bs}$ tends to the infinite length value.

In the regime ($u \gg 1$), the current $I_{bs}$ also depends on the dimensionless parameters $\omega_{a} = |a|\Omega /v$ and $\omega_{r} = |r|\Omega /v$, that characterize the role of the geometry, i.e.\ the relation between the relative ($a$) and medium position ($r$) of the impurities with the dynamic length scale $v/\Omega$.  The dependence of $I_{bs}$ with $r$, that as expected was not present in the case of $L = \infty$, is due to the breaking of the translational invariance.  When $\omega_{a,r} \ll 1$ the dependence with the position of the impurities in the relation between the pumping current at finite and infinite length can be dropped.
In the opposite limit, when $u\gg\omega_{a,r}\gg 1$, the current becomes a superposition of damped oscillatory functions of  $\omega_{r}$ and $\omega_{a}$, both with period $2 \pi$ and a decaying factor of the form $\omega_{a}^{- K \gamma^{2n}}$ or  $\omega_{r}^{- K \gamma^{2n + 1}}$.

Concerning the small length regime ($u \ll 1$), the whole structure of damped oscillatory functions
disappears and $I_{bs}$ is proportional to $\Omega$ and $L^{2 - 2K}$, showing a more pronounced suppression of the current for strong interactions.  In this case the dependence with the geometry is modulated by a factor which is maximum when $|a|/L$ and $|r|/L$
tend to zero.

\acknowledgments This work was partially supported by Universidad Nacional de La
Plata (Argentina) and Consejo Nacional de Investigaciones Cient\'
ificas y T\'ecnicas, CONICET (Argentina). SFV was partially supported by the European Comission through an Eurotango scholarship.

\appendix

\section{Computation of exponentials of the $\Phi$-fields at finite length}

In this Appendix we compute the expectation value in (\ref{factorA}) for finite length and zero temperature; this leads to expression (\ref{pump6}) for the backscattered current $I_{bs}$. We begin by considering a scalar field $\widehat{\Phi}(x, t)$ in a finite one-dimensional spacelike region of size $D$; we will impose periodic boundary conditions at $x=\pm D/2$ and 
take the limit $D\rightarrow \infty$. The field satisfies the following equation of motion derived from the Lagrangian density (\ref{L0})
\begin{equation}\label{eqmo}
\left(\frac{\partial^2}{\partial t^2}-\frac{\partial}{\partial x}v^2(x)\frac{\partial}{\partial x}\right)\,\widehat{\Phi}(x, t)=0\,.
\end{equation}

The solutions of (\ref{eqmo}) can be written as
\begin{equation}\label{field}
 \hat{\Phi}(x,t)=\sum_{p>0}\hat{a}_p\,e^{-i\,p\, v_Ft}\,\phi_{p}(x)+\hat{a}_p^{\dagger}\,e^{i\,p\,v_Ft}\,\phi_{p}^{*}(x)\,,
\end{equation}
where $a_p^\dagger$ and $a_p$ satisfy the algebra of creation and annihilation operators and $\phi_p(x)$ are given by\begin{eqnarray}
    \phi_p^+=C_+\,\cos{(Kpx)}\,,\\
    \phi_p^-=C_-\,\sin{(Kpx)}\,.
\end{eqnarray}

The quantized momenta $p$ are the positive solutions of the 
equation\begin{multline}\label{spec}
f_\pm(p) = (1-K)\,\sin\left\{\frac{p}{2}\left[(1+K)L-D\right]\right\}\\\mp(1+K)\,\sin\left\{\frac{p}{2}\left[(1-K)L-D\right]\right\}=0\,,
\end{multline}where the upper (lower) sign corresponds to $\phi^+_p$ ($\phi^-_p$). To leading order in $D$ the normalization constants $C_\pm$ take the following values:\begin{equation}
    C_{\pm}=\frac{2K}{\sqrt{D}}\left[1+K^2\mp(1-K^2)\cos{(KpL)}\right]^{-1/2}\,.
\end{equation}

From (\ref{field}) we can express the v.e.v.\ in (\ref{factorA}) in terms of an infinite sum over all positive solutions $p$ of (\ref{spec}). This sum can be written as an integral in the complex $p$-plane along a contour which encloses the quantized momenta $p$ if we introduce in the integrand the logarithmic derivative $\partial_p \ln{f_{\pm}(p)}$. Afterwards, the contour of integration can be deformed to the positive real semi-axis. The result, for large $D$, reads\begin{eqnarray}\label{integral}
\langle0|\hat{\Phi}(x',t')\hat{\Phi}(x,t)-\frac{\hat{\Phi}^2(x,t)+\hat{\Phi}^2(x',t')}{2}|0\rangle=\frac{D}{4\pi v_F}\times\nonumber\\\mbox{}\times
\sum_{s=\pm} \int_{0}^{\infty} \frac{dp}{p}\,e^{-\Lambda\,p}\big[\phi_{p}^s(x)\phi_{p}^s\mbox{}^{*}(x')\,e^{-i\,v_F\,p\,(t - t')}- \nonumber\\ \mbox{}-\frac{1}{2}\phi_{p}^s(x)\phi_{p}^s\mbox{}^{*}(x)-\frac{1}{2}\phi_{p}^s(x')\phi_{p}^s\mbox{}^{*}(x')\big]\,,\nonumber\\
\end{eqnarray}where $\Lambda$ is a small length regulator. From (\ref{integral}) we obtain
\begin{multline}\label{a11}
A_{1,1} = \exp[ - K \ln \{ \frac{ (\Lambda - i v(t - t') )^{2} + (x - x')^{2}} {\Lambda^{2}}\} \\  - K\sum_{n \in Z even \neq 0} \gamma^{|n|}\ln \{ \frac{ ((\Lambda/L) - i v(t - t')/L )^{2} + (n + \frac{x - x'}{L})^{2}} {n^{2}} \} \\ - K\sum_{n \in Z odd } \gamma^{|n|}\ln \{ \frac{ ((\Lambda/L) - i v(t - t')/L )^{2} + (n + \frac{x + x'}{L})^{2}} {(n + 2x/L )(n + 2x'/L) } \} ].
\end{multline}

We remark that the logarithms in (\ref{a11}) are defined with their branch cuts on the negative real semi-axis.


\begin{thebibliography}{99}

\bibitem{Chamon0} J.E. Moore, P. Sharma and C. Chamon, Phys. Rev. B {\bf 62}, 7298 (2000).

\bibitem{Giamarchi}T. Giamarchi, Quantum Physics in One dimension, (Clarendon Press,
Oxford, 2004).

\bibitem{Ventra}M. Di Ventra, Electrical Transport in Nanoscale Systems, (Cambridge
University Press, 2008).

\bibitem{Nazarov}Y. V. Nazarov, and Y. M. Blanter, Quantum Transport, (Cambridge
University Press, 2009).

\bibitem{Gogolin1} A.O. Gogolin Phys. Rev. Lett. {\bf 71}, 2995 (1993).

\bibitem{Chamon2} P. Sharma and C. Chamon, Phys. Rev. Lett. {\bf 87}, 096401 (2001).

\bibitem{Nature} M.J. Baird, F.R. Hope and A.F.G Wyatt, Nature {\bf 304}, 325 (1983).

\bibitem{Science} Philippe Poncharal, Z. L. Wang, Daniel Ugarte, and Walt A. de Heer, Science {\bf 283}, 1513 (1999).

\bibitem{Science2} Ray H. Baughman, Changxing Cui, Anvar A. Zakhidov, Zafar Iqbal, Joseph N. Barisci, Geoff M. Spinks, Gordon G. Wallace, Alberto Mazzoldi, Danilo De Rossi, Andrew G. Rinzler, Oliver Jaschinski, Siegmar Roth, and Miklos Kertesz, Science {\bf 284}, 1340 (1999).

\bibitem{Miliken} F.P. Miliken, C.P. Umbach and R.A. Webb, Solid State Commun. {\bf 97}, 309 (1996).

\bibitem{Fujisawa} T. Fujisawa, T. Hayashi and S. Sasaki, Rep. Prog. Phys. {\bf 69}, 759 (2006).

\bibitem{Torres} L.E. Foa Torres and G. Cuniberti, C.R. Physique. {\bf 10}, 297 (2009).

\bibitem{Chamon 1} P. Sharma and  and C. Chamon, Phys. Rev. B {\bf
68}, 035321 (2003).

\bibitem{Feldman} D. E. Feldman and Y. Gefen, Phys. Rev. B {\bf 67},
115337 (2003).

\bibitem{Schmeltzer} D. Schmeltzer, Phys. Rev. B {\bf 63},
125332 (2001).

\bibitem{yo 1} M.J. Salvay, H.A. Aita and C.M.  Na\'on, Phys. Rev. B {\bf 81}, 125406
(2010).

\bibitem{Perfetto} E. Perfetto, G. Stefanucci and M. Cini, Phys. Rev. Lett. {\bf 105}, 156802 (2010).

\bibitem{pump 0} Complete reviews on pumping mechanisms are: G. Platero and R. Aguado, Phys. Rep. {\bf 395}, 1 (2004);  S. Kohler, J. Lehmann and P.
H\"anggi, Phys. Rep. {\bf 406}, 379 (2005); P.
H\"anggi and F. Marchesoni, Rev. Mod. Phys. {\bf 81}, 387 (2009).

\bibitem{exp 2}M. Switkes, C. M. Marcus, K. Campman, and A. C. Gossard,
Science {\bf 283}, 1905 (1999).

\bibitem{exp 3} L. J. Geerligs, V. F. Anderegg, P. A. M. Holweg, J. E. Mooij, H.
Pothier, D. Esteve, C. Urbina, and M. H. Devoret, Phys. Rev.
Lett. {\bf 64}, 2691 (1990).

\bibitem{exp 4} L. DiCarlo, C. M. Marcus, and J. S. Harris, Phys. Rev. Lett. {\bf 91},
246804 (2003).

\bibitem{exp 5}M. G. Vavilov, L. DiCarlo, and C. M. Marcus, Phys. Rev. B {\bf 71},
241309(R) (2005).

\bibitem{exp 6}P. J. Leek, M. R. Buitelaar, V. I. Talyanskii, C. G. Smith, D.
Anderson, G. A. C. Jones, J. Wei, and D. H. Cobden, Phys. Rev.
Lett. {\bf 95}, 256802 (2005).

\bibitem{pump 2} L. Arrachea, Phys. Rev. B {\bf
72}, 121306(R) (2005); {\bf
72}, 249904(E) (2005).

\bibitem{pump 3} L. Arrachea, C. Na\'on and M. Salvay, Phys. Rev. B {\bf
76}, 165401 (2007).

\bibitem{pump 5} A. Soori and D. Sen, Phys. Rev. B {\bf
82}, 115432 (2010).

\bibitem{pump 6} A. Saha and S. Das, Phys. Rev. B {\bf
78}, 075412 (2008).

\bibitem{pump 7} M. Moskalets and M. Buttiker,  Phys. Rev. B {\bf
78}, 035301 (2008); {\bf
68}, 161311(R) (2003); {\bf
66}, 205320 (2002); {\bf
66}, 035306 (2002).

\bibitem{pump 9} P.W. Brouwer, Phys. Rev. B {\bf
58}, 10135 (1998).

\bibitem{pump 10} M. L. Polianski and P.W. Brouwer, Phys. Rev. B {\bf
84}, 075304 (2001).

\bibitem{pump 11} M. Strass, P.
H\"anggi and S. Kohler, Phys. Rev. Lett. {\bf
95}, 130601 (2005).

\bibitem{pump 4} A. Kundu, S. Rao and A. Saha, Phys. Rev. B {\bf
83}, 165451 (2011).

\bibitem{pump 8} P. San-Jose, E. Prada, S. Kohler and H. Schomerus, Phys. Rev. B {\bf
84}, 155408 (2011).

\bibitem{theories} A. Agarwal and D. Sen, Phys. Rev. B {\bf
76}, 035308 (2007).

\bibitem{yo 2} M.J. Salvay, Phys. Rev. B {\bf
79}, 235405 (2009).

\bibitem{Dolcini} F. Dolcini, H.
Grabert, I. Safi and B. Trauzettel  Phys. Rev. Lett. {\bf 91}, 266402 (2003).

\bibitem{Dolcini2} F. Dolcini, B. Trauzettel, I. Safi and H.
Grabert, Phys. Rev. B {\bf 71}, 165309 (2005).

\bibitem{yo 3} M. Salvay, A. Iucci and C.  Na\'on, Phys. Rev. B {\bf 84}, 075482
(2011).

\bibitem{Cheng} F. Cheng and G. Zhou, Phys. Rev. B {\bf 73}, 125335
(2006).

\bibitem{Keldysh} J. Schwinger, J. Math. Phys. {\bf 2}, 407 (1961);
L.V. Keldysh, Soviet Physics JETP {\bf 20}, 1018 (1965);  E. M. Lifshitz, and L. P. Pitaevskii, Statistical
Physics, Part II, (Pergamon Press, Oxford, 1980); G. D. Mahan, Many-Particle Physics, Third Edition (Kluwer
Academics/Plenum Publishers, 2000); A. Das, Finite Temperature Field Theory, (World Scientific, 1997); A. Kamenev and A. Levchenko, Advances in Physics {\bf 158} No. 3,
197 (2009).

\bibitem{GGM}D. B. Gutman, Yuval Gefen, and A. D. Mirlin,
Europhysics Letters {\bf 90}, 37003 (2010) , Phys. Rev. B {\bf 81}, 085436 (2010) .

\bibitem{standardbos} M. Stone, Bosonization, (World Scientific,
1994); A. O. Gogolin, A. A. Nersesyan, and A. M. Tsvelik, Bosonization in
strongly correlated systems, (University Press, Cambridge, 1998).

\bibitem{makogon} D. Makogon, V. Juricic and C.M. Smith, Phys. Rev. B {\bf 74}, 165334
(2006).

\bibitem{ponomarenko} V.V. Ponomarenko and N. Nagaosa, Phys. Rev. B {\bf
56}, R12756 (1997).



\end{thebibliography}
\end{document}